\documentclass[12pt]{article}
\usepackage{subfloat}
\pdfoutput = 1
\usepackage{subcaption}
\usepackage{hyperref}
\usepackage{graphicx} 
\usepackage{amsmath}
\usepackage{amsfonts}   
\usepackage{amssymb}
\usepackage{multirow}
\usepackage{hhline}
\begin{document}
\date{Today}
\title{{\bf{\Large  Rainbow black hole thermodynamics and the generalized uncertainty principle }}}

\author{
{\bf {\normalsize Rituparna Mandal}$^{a}$
\thanks{drimit.ritu@gmail.com, rituparna1992@bose.res.in}},\,
{\bf {\normalsize Sukanta Bhattacharyya}$^{b}$
\thanks{sukanta706@gmail.com}},\,
{\bf {\normalsize Sunandan Gangopadhyay}$^{a}$\thanks{sunandan.gangopadhyay@gmail.com, sunandan.gangopadhyay@bose.res.in}}\\
$^{a}$ {\normalsize Department of Theoretical Sciences,}\\{\normalsize S.N. Bose National Centre for Basic Sciences,}\\{\normalsize JD Block, 
Sector III, Salt Lake, Kolkata 700106, India}\\[0.1cm]
$^{b}$ {\normalsize Department of Physics, West Bengal State University, Barasat, Kolkata 700126, India}\\[0.1cm]}
\date{}

\maketitle

\begin{abstract}
\noindent We study the phase transition of rainbow inspired higher dimensional Schwarzschild black hole incorporating the effects of the generalized uncertainty principle. First, we obtain the relation between the mass and Hawking temperature of the rainbow inspired black hole taking into account the effects of the modified dispersion relation and the  generalized uncertainty principle. The heat capacity is then computed from this relation which reveals that there are remnants. The entropy of the black hole is next obtained in $3+1$ and $4+1$-dimensions and is found to have logarithmic corrections only in $3+1$-dimensions. We further investigate the local temperature, free energy and stability of the black hole in an isothermal cavity. From the analysis of the free energy, we find that there are two Hawking-Page type phase transitions in $3+1$ and $4+1$-dimensions if we take into account the generalized uncertainty principle. However, in the absence of the generalized uncertainty principle, only one Hawking-Page type phase transition exists in spacetime dimensions greater than four.
  
\end{abstract}

\vskip 1cm

\section{Introduction}
All quantum gravity (QG) theories namely, loop quantum gravity \cite{rc,ca}, string theory \cite{aci}, noncommutative geometry \cite{gf}, indicates that there exists an observer independent minimum length scale which can be identified with the Planck length \cite{ma,pmi}. However, the feature of observer independence of the minimum length does not agree with the Lorentz transformations in special theory of relativity (STR) since the Planck length scale is not a Lorentz invariant quantity. To resolve this contradiction, the standard dispersion relation in STR, has been modified. The modified dispersion relation (MDR) appears in a new theory known as a doubly special relativity (DSR) \cite{ac}-\cite{ac1}. This theory has two invariants namely, the speed of light $c$ and the Planck length $l_{p}$ (or the Planck energy $E_{p}$). Based on this platform, rainbow gravity (RG) was introduced in \cite{ms}. It was proposed here that the energy of the test particle can also affect the geometry of the spacetime and hence the name gravity's rainbow. The spacetime can no longer be described by a single metric, but by a family of metrics (rainbow of metrics), parameterized by the ratio $(E/E_{p})$. The deformed energy-momentum relation of the test particle in the RG background takes the form \cite{ms,mjs}
\begin{equation}
 E^{2}f^{2}(E/E_{p})-p^{2}g^{2}(E/E_{p})=m^{2}
\label{eq3}
\end{equation}
where $m$, $p$ are the mass and the momentum of the test particle, $f(E/E_{p})$ and $g(E/E_{p})$ are known as the rainbow functions. These functions get constrained to reproduce the standard dispersion relation in the infrared limit, that is, $\lim\limits_{E/E_p\rightarrow 0}f\left(E/E_p\right) = 1$; $\lim\limits_{E/E_p\rightarrow 0}g\left(E/E_p\right)=1$. Loop quantum gravity considerations lead to the forms for these rainbow functions \cite{amu}-\cite{sl}
\begin{eqnarray}
f\left(E/E _{p}\right)=1~~;~~~~ g\left(E/E _{p}\right)=\sqrt{1-\eta (E/E_{P})^{n}}
\label{1}
\end{eqnarray} 
where $\eta$ is the rainbow parameter\footnote{Note that loop quantum gravity does not impose any restriction on the value of $n$.}.

\noindent The existence of a minimum length scale has also be taken into account in the uncertainty principle. This has led to modifying the Heisenberg uncertainty principle (HUP) to the generalized uncertainty principle (GUP) \cite{kem}-\cite{chn}. This idea can have important consequences in various area of physics. One of the most interesting area where the GUP has played an important role is the study of black hole thermodynamics \cite{mkp}-\cite{adler}. It has been understood that the inclusion of the effects of the GUP in the physics of black holes gives rise to the existence of black hole remnants \cite{adler}-\cite{sa1}. Further, the entropy of the black hole also gets logarithmic corrections to the famous area law. It has been realized that the presence of remnants in the rainbow inspired black holes using the standard dispersion relation could resolve the catastrophic behaviour of the Hawking radiation as the black hole mass shrinks to zero \cite{Aali}-\cite{sa}. To place our work in the proper context, we would like to mention that  there has been a lot of work on black hole thermodynamics in the rainbow inspired black holes in the last few years \cite{lin}-\cite{li}. 

\noindent In this paper, we investigate the thermodynamics of a rainbow-inspired higher dimensional Schwarzschild black hole in the MDR platform using the simplest form of GUP \cite{adler}. First we explore the modification of the thermodynamic properties, namely, the Hawking temperature, heat capacity and entropy of the higher dimensional Schwarzschild black hole due to the presence of both the GUP and MDR in the RG background. We start by obtaining a relation between the mass and Hawking temperature of the rainbow inspired black hole. This prepares us to investigate whether there are remnants in this set up. Interestingly, we find that there are no remnants in the RG set up if MDR is taken into account in the absence of the GUP. However, remnants are present in the RG background if the usual dispersion relation is considered in the absence of the GUP. Then we compute the Hawking temperature, specific heat and entropy for $d=4$ and $d=5$ spacetime dimensions. In the next part of our work, we investigate the phase transition and the thermodynamic stability of the RG inspired black hole. To do this, we consider the black hole inside a finite spherical concentric cavity whose radius is larger than the horizon radius of the black hole. We first calculate the on-shell free energy of the higher dimensional Schwarzschild black hole in RG incorporating the effects of the standard form of the GUP. Doing this helps us to observe the Hawking-Page type phase transition for $d=4,5$ for both the cases $\alpha=0$ and $\alpha \neq 0 $, where $\alpha$ is the GUP parameter.

\noindent The paper is organized as follows. In section 2, we study the modification of the thermodynamic properties of the higher dimensional RG inspired Schwarzschild black hole incorporating the GUP and using the MDR. In section 3, we investigate the phase transition and thermodynamic stability of the black hole. We conclude in section 4.

\section{Heat capacity and entropy of the black hole}
In this paper we want to explore the thermodynamic properties of the higher dimensional Schwarzschild black hole in the RG background with the specific choice of rainbow functions \eqref{1} incorporating the effects of the GUP. 

\noindent The metric of the RG inspired Schwarzschild black hole in $d$-dimensions reads \cite{af}
\begin{equation}
ds^{2}=-\frac{1}{f^{2}({E/E_{p}})}\left (1-\frac{\mu}{r^{d-3}}\right)dt^{2} + \frac{1}{g^{2}(E/E _{p})} \left(1-\frac{\mu}{r^{d-3}} \right)^{-1} dr^{2} + \frac{r^{2}}{g^{2}(E/E _{p})} d\Omega^{2}_{d-2}
\label{eq6}
\end{equation}
where the constant $ \mu $ has the form
\begin{equation}
\mu=\frac{16 \pi G_{d}M}{(d-2)\Omega_{d-2}}
\label{eq7}
\end{equation}
and $ \Omega_{d-2} $ is the volume of the $(d-2)$-sphere given as
\begin{equation}
\Omega_{d-2}= \frac{2 \pi^{\frac{d-1}{2}}}{\Gamma(\frac{d-1}{2})}~~.
\label{eq8}
\end{equation}
The horizon radius $r_{+}$ of the above black hole can be obtained by solving $(1-\frac{\mu}{r_{+}^{d-3}})=0$. This yields 
\begin{eqnarray}
r_{+}=\mu^{\frac{1}{d-3}}=\frac{1}{\sqrt{\pi}}\left (\frac{8M\Gamma(\frac{d-1}{2})}{E_{p}^{d-2}(d-2)} \right )^{\frac{1}{d-3}}~.
\label{eq9}
\end{eqnarray} 
\noindent Now for any massless quantum particle near the horizon of a black hole, the energy of the particle can be obtained by employing the MDR (\ref{eq3}) and is given by
\begin{eqnarray}
E=\frac{p g\left(E/E_{p}\right)}{f \left(E/E _{p}\right)}~.
\label{2}
\end{eqnarray}
This in turns characterizes its temperature as
\begin{eqnarray}
T = \frac{1}{k_B} \frac{p g(E/E_p)}{f(E/E_p)}
\label{111}
\end{eqnarray}
where $k_B$ is the Boltzmann constant. This temperature of the particle will be the same as the temperature of the black hole in thermal equilibrium. Setting $k_B=1$ and using the form of the rainbow functions (\ref{1}), the above relation for the temperature of the black hole takes the form 
\begin{eqnarray}
T = p \sqrt{1-\eta (E/E_{P})^{n}}~.
\label{eq10}
\end{eqnarray}
\noindent Our next step is to obtain a relation between the temperature of the black hole $(T)$ and the black hole mass $(M)$. To do this, we take recourse to the GUP \cite{adler}
\begin{equation}
\Delta x \Delta p\geq \frac{1}{2}\left [ 1 + \alpha l_{p}^{2}(\Delta p)^{2} \right ]
\label{eq1}
\end{equation}
\noindent where $l_{p}=\frac{1}{E_{p}}=G_{d}^{\frac{1}{(d-2)}}$ is the Planck length, $G_{d}$ is the universal Newton's gravitational constant in $ d $-dimensions, $E_{p}$ is the Planck energy and $\alpha$ is a dimensionless real constant of order unity and we have set $\hbar=1$. It is now expected that near the horizon of the rainbow inspired $d$-dimensional Schwarzschild black hole, the position uncertainty is of the order of the horizon radius $r_+$, namely,

\begin{eqnarray}
\Delta x = \epsilon r_+
\label{9}
\end{eqnarray}
where $\epsilon$ is a calibration factor and the momentum uncertainty $\Delta p \sim p$. Substituting $\Delta x$ from eq.(\ref{9}) and $\Delta p \sim p$  in the saturated form of the GUP (\ref{eq1}), and using eq.(\ref{eq10}), we obtain the relationship between the horizon radius (and hence mass of the black hole) and the temperature of the black hole. This reads
\begin{eqnarray}
r_+=\frac{1}{2 \epsilon \sqrt{1-\frac{\eta {T}^2}{T_p^{2}}}}\left[\frac{1}{T}(1-\frac{\eta T^2}{T_p^{2}})+\frac{\alpha T}{T_p^{2}}\right]~.
\label{00}
\end{eqnarray}
To obtain the calibration  factor $\epsilon$, we note that the Hawking temperature of the Schwarzschild black hole in $d$-dimensions (in the absence of the rainbow parameter $\eta$ and the GUP parameter $\alpha$) is given by
\begin{eqnarray}
T = \frac{(d-3)}{4 \pi r_{+}}~~.
\label{10}
\end{eqnarray}
This fixes $\epsilon$ to be
\begin{eqnarray}
\epsilon = \frac{2 \pi}{(d-3)}~.
\label{11}
\end{eqnarray}
Hence eq.(\ref{00}) now takes the form 
\begin{eqnarray}
r_+ =  \frac{(d-3)}{ 4 \pi \sqrt{1-\frac{\eta {T}^2}{T_p^{2}}}}\left[\frac{1}{T}(1-\frac{\eta T^2}{T_p^{2}})+\frac{\alpha T}{T_p^{2}}\right]~.
\label{12}
\end{eqnarray}
From the above expression, we observe that $(1-\frac{\eta T^2}{T_p^{2}}) \geq 0$  for the reality of $r_+$. This in turn implies that  $T \leq \frac{T_p}{\sqrt{\eta}}$. Interestingly, this upper bound on the temperature $T$ holds in all dimensions. Further, the upper bound on the temperature owes its origin to RG. In the limit, $\eta \rightarrow 0$, there is no upper bound on the temperature.

\noindent Now the heat capacity of the black hole can be defined as
\begin{eqnarray}
C=\frac{dM}{dT}= \left(\frac{dM}{dr_{+}}\right) \left(\frac{dr+}{dT}\right)~~.
\label{01}
\end{eqnarray}
Using eq.(s) (\ref{eq9}) and (\ref{00}), the heat capacity in terms of the temperature reads
\begin{eqnarray}
C=-\frac{2(d-2)T_p^{d-2}\pi^{\frac{d+1}{2}} T (1-\frac{T^{2}(\alpha +\eta)}{T_p^{2}})\left(\frac{(d-3)(1+\frac{T^{2}(\alpha -\eta)}{T_p^{2}})}{4\pi T \sqrt{(1-\frac{T^{2}\eta}{T_p^{2}})}}\right)^{(d-1)}}{(d-3)(1+\frac{T^{2}(\alpha -\eta)}{T_p^{2}})^3\Gamma[\frac{(d-1)}{2}]}~~.
\label{02}
\end{eqnarray}
Now to obtain the remnant mass at which the radiation process terminates, we set $C=0$. This first gives the temperature of the black hole (at which the radiation process stops) to be
\begin{eqnarray}
T_{rem} = \frac{T_p}{\sqrt{\alpha + \eta}}~~.
\label{03}
\end{eqnarray}
Using this in eq.(\ref{00}), we get the horizon radius corresponding to this temperature to be  
\begin{eqnarray}
(r_{+})_{rem}=\frac{(d-3)\sqrt{\alpha}}{2\pi T_p}~~.
\label{04}
\end{eqnarray}
The above expression yields the remnant mass for the RG inspired Schwarzschild black hole in $d$-dimensions :
\begin{eqnarray}
M_{rem}=\frac{(d-2)T_{p}^{d-2}}{8\Gamma\left(\frac{d-1}{2}\right)}\left(\frac{(d-3)\sqrt{\alpha \pi}}{2\pi T_{p}}\right)^{\frac{1}{(d-3)}}~.
\label{m}
\end{eqnarray}
The dependency of the remnant mass on the dimension of the spacetime is seen from the above expression. It is interesting to note that the remnant mass does not depend on the rainbow parameter $\eta$ and depends only on the GUP parameter $\alpha$. For $\alpha = 0$, we observe that there exists no remnants which is consistent with earlier findings in the literature \cite{adler}-\cite{pc}. Further, setting $d=4$, we find that
\begin{eqnarray}
M_{rem} = \frac{\sqrt{\alpha}}{4 \pi} T_{p}
\label{011}
\end{eqnarray}
which agrees with that obtained in \cite{sg}.

\noindent We now proceed to obtain the entropy of the black hole from the first law of black hole thermodynamics which reads
\begin{equation}
S = \int \frac{dM}{T}~.
\label{012}
\end{equation}
To evaluate this integral, we need to obtain a relation between the temperature $T$ and mass of the black hole $M$. To do this, we first look at the GUP (\ref{eq1}) and recast it in the form
\begin{eqnarray}
p \approx \Delta p =\frac{\Delta x}{ \alpha l_{p}^{2}}{\left [ 1-\sqrt{1-\frac{\alpha l_{p}^{2}}{(\Delta x)^{2}}} \right ]}
\label{eq2}
\end{eqnarray}
where the negative sign before the square root has to taken in order to recover the HUP in the limit $\alpha \rightarrow 0$. It is to be noted that since the uncertainty in the momentum must be real, hence the uncertainty in position $\Delta x \geq \sqrt{\alpha}l_{p}$. Substituting eq.(s) (\ref{9}, \ref{11}, \ref{eq2}) in eq.(\ref{eq1}), we get the expression for the temperature in terms of the mass as 
\begin{eqnarray}
T=\frac{2 \pi r_{+}}{(d-3)\alpha l_{p}^{2}}{\left [ 1 - \sqrt{1-\frac{(d-3)^2 \alpha l_{p}^{2}}{ 4 \pi^2 r_{+}^{2}}} \right ]}\sqrt{1-\eta (E/E_{P})^{n}}~.
\label{eq11}
\end{eqnarray}

\begin{figure}
\begin{subfigure}{.5\textwidth}
  \centering
    \includegraphics[width = 7.5 cm, height = 5.5cm]{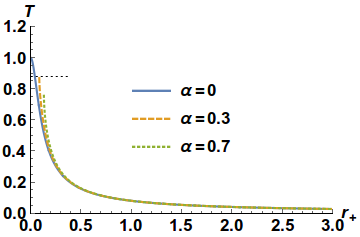}
  \caption{d=4}
  \label{fig:sfig1}
\end{subfigure}%
\begin{subfigure}{.5\textwidth}
  \centering
  \includegraphics[width= 7.5 cm, height = 5.5cm]{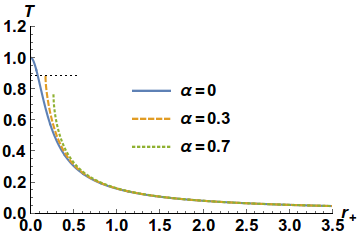}
  \caption{d=5}
  \label{fig:sfig2}
\end{subfigure}%
\caption{The modified Hawking temperature $T$ vs. event horizon radius $r_{+}$ for $d=4,5$. The rainbow parameter $\eta=1$ and the GUP parameter $\alpha=0, 0.3, 0.7$ ($l_{p}=\frac{1}{E_{p}}=1$).}
\label{fig:fig1}
\end{figure}

\noindent It is reassuring to note that in the limit $\eta \rightarrow 0$, the above expression for the temperature takes the form
\begin{eqnarray}
T_{G}=\frac{2 \pi r_{+}}{(d-3)\alpha l_{p}^{2}}{\left [ 1 - \sqrt{1-\frac{(d-3)^2 \alpha l_{p}^{2}}{ 4 \pi^2 r_{+}^{2}}} \right ]}
\label{eq12}
\end{eqnarray}
which for $d = 4$ reduces to 
\begin{eqnarray}
T = \frac{2 \pi r_{+}}{\alpha l_{p}^2} \left[ 1 - \sqrt{1 - \frac{\alpha l_{p}^2}{4 \pi^2 r_{+}^2}} \right]~.
\label{A}
\end{eqnarray}
This agrees with that obtained in \cite{adler}, \cite{sg}.

\noindent Looking back at eq.(\ref{eq11}), we now proceed to calculate $E/E_{p}$ for a massless particle. This can be done by substituting $p$ from eq.(\ref{eq2}) in the MDR (\ref{eq3}) (with $m = 0$). For $n = 2$, we obtain
\begin{eqnarray}
\frac{E}{E_{p}}=\frac{\frac{2 \pi  r_{+} E_{p}}{(d-3) \alpha }\left ( 1-\sqrt{1-\frac{(d-3)^2 \alpha}{4 \pi^2 r_{+}^{2} E_{p}^{2}}} \right )}{\sqrt{1 + \frac{ 4 \eta \pi^2 r_{+}^{2}E_{p}^{2} }{(d-3)^2 \alpha ^{2}}\left( 1-\sqrt{1-\frac{(d-3)^2 \alpha}{4 \pi^2 r_{+}^{2}E_{p}^{2}}}\right )^{2}}}~~.
\label{eq13}
\end{eqnarray}
The above relation for the energy of the massless particle is used in eq.(\ref{eq11}) to get the temperature of the black hole which incorporates the effects of RG and the GUP
\begin{eqnarray}
T = \frac{\frac{2 \pi r_{+} E_p^2}{ (d-3) \alpha}\left ( 1 - \sqrt{1 - \frac{(d-3)^2 \alpha }{4 \pi^2 r_{+}^{2} E_{p}^{2}}} \right )}{\sqrt{1 + \frac{4 \pi^2 r_{+}^{2}E_{p}^{2}\eta }{(d-3)^2 \alpha ^{2}}\left( 1 - \sqrt{1 - \frac{(d-3)^2 \alpha}{4 \pi^2 r_{+}^{2}E_{p}^{2}}}\right )^{2}}}~~.
\label{eq14}
\end{eqnarray}
In Fig.\ref{fig:fig1}, we plot this temperature with the radius of the event horizon of the black hole in $d=4$ and $5$ dimensions for different values of the GUP parameter $\alpha$ and setting the rainbow parameter $\eta$ to unity. From the plot, we can observe that for $\alpha=0$, the temperature is finite at $r_{+}=0$. Hence the temperature gets regularized in the presence of RG. In the presence of the GUP, that is for $\alpha \neq 0$, the temperature is finite at a non-zero value of the event horizon and therefore shows the existence of remnants. 

\noindent As a consistency check, we note that if we substitute $(r_{+})_{rem}$ from eq.(\ref{04}) in the above equation, we recover eq.(\ref{03}). Further, in the limit  $\alpha \rightarrow 0$, we obtain 
\begin{eqnarray}
T=\frac{(d-3) E_p}{\sqrt{(16 \pi^2 E_{p}^{2}r_{+}^{2}+ (d-3)^2\eta )}}~~.
\label{eq15}
\end{eqnarray}
This result for the temperature holds for the rainbow inspired Schwarzschild  black hole in $d$-dimensions. In $d= 4$, we get \footnote{It should be noted that the expression agrees upto a factor with \cite{wk}.} 
\begin{eqnarray}
T=\frac{E_p}{\sqrt{(16 \pi^2 E_{p}^{2}r_{+}^{2} + \eta )}}~~.
\label{B}
\end{eqnarray}

\begin{figure}
\begin{subfigure}{.5\textwidth}
  \centering
  \includegraphics[width=  7cm, height =5 cm]{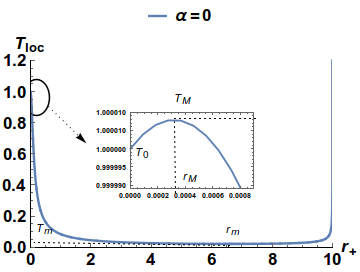}
  \caption{d=4}
  \label{fig:sfig3}
\end{subfigure}
\begin{subfigure}{.5\textwidth}
	\centering
	\includegraphics[width= 7cm, height = 5 cm]{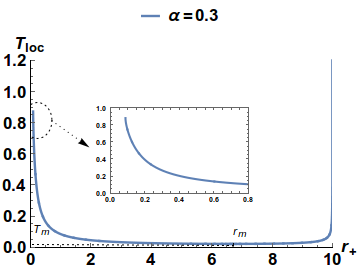}
	\caption{d=4}
	\label{fig:sfig4}
\end{subfigure} \nonumber\\
\begin{subfigure}{.5\textwidth}
  \centering
  \includegraphics[width= 7cm, height = 5 cm]{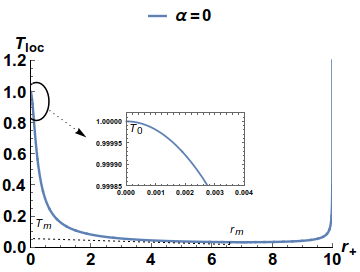}
  \caption{d=5}
  \label{fig:sfig5}
\end{subfigure}
\begin{subfigure}{.5\textwidth}
\centering
	\includegraphics[width= 7cm, height = 5 cm]{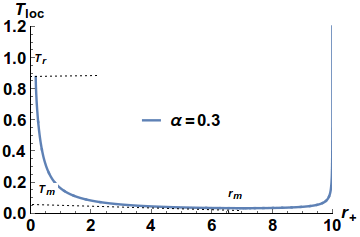}
	\caption{d=5}
	\label{fig:sfig6}
\end{subfigure}
\caption{Local temperature $T_{loc}$ vs. horizon  radius $r_{+}$ of the black hole in $d=4,5$. Rainbow parameter $\eta=1$,~$r=10$~.}
\label{fig:fig2}
\end{figure}

\noindent Using the expression for the temperature of the black hole (\ref{eq14}) in eq.(\ref{012}), we obtain
\begin{eqnarray}
S &=&\frac{(d-2)(d-3)^{2}E_{p}^{(d-4)}\pi^(\frac{d-3}{2})}{16 \pi \Gamma (\frac{d-1}{2})}\int \frac{r_{+}^{(d-5)} \alpha \sqrt{1 + \frac{4 \pi^2 r_{+}^{2}E_{p}^{2}\eta }{(d-3)^2 \alpha ^{2}}\left( 1 - \sqrt{1 - \frac{(d-3)^2 \alpha}{4 \pi^2 r_{+}^{2}E_{p}^{2}}}\right )^{2}}dr_{+}} {\left ( 1 - \sqrt{1 - \frac{(d-3)^2 \alpha }{4 \pi^2 r_{+}^{2} E_{p}^{2}}} \right )} \nonumber \\ &  = & \frac{(d-2)(d-3)E_{p}^{(d-6)} \pi^{(\frac{(d-3)}{2})}}{512 \epsilon^3\Gamma (\frac{d-1}{2})} \int r_{+}^{(d-7)}  \left ( 128 \epsilon^4 r_+^4 E_p^4 - 32 \epsilon^2 r_+^2 E_p^2 \alpha  + 16 \epsilon^2 r_+^2 E_p^2 \eta \right. \nonumber \\ && \left. - 8 \alpha^2 + 4 \eta \alpha -\eta^2 \right )dr_{+} + \mathcal{O}(\alpha^{3},\eta \alpha^{2}, \eta^2 \alpha, \eta^{3})~. 
\label{eq17}
\end{eqnarray}
For $d=4$, the above expression takes the form  
\begin{eqnarray}
S & = &  \pi E_{p}^{2}r_{+}^{2}-(\frac{\alpha}{8\pi}-\frac{\eta}{16\pi}) \ln(r_{+}E_{p})+\frac{\alpha^{2}}{256 \pi^{3} E_{p}^{2}r_{+}^{2}}-\frac{\alpha \eta}{512 \pi^{3} E_{p}^{2}r_{+}^{2}} +\frac{\eta^{2}}{2048 \pi^{3} E_{p}^{2}r_{+}^{2}} \nonumber \\ && + S_{0} + \mathcal{O}(\alpha^{3},\eta \alpha^{2}, \eta^2 \alpha, \eta^{3})
\label{eq18}
\end{eqnarray}
where $S_{0}$ is the constant of integration. We can determine the integration constant $S_{0}$ by requiring that $S\rightarrow 0$ (since $C\rightarrow 0 $) as $r_{+} \rightarrow (r_{+})_{rem}$. This gives using eq.(\ref{04}) 
\begin{eqnarray}
S_{0} =- \left[ (\frac{\alpha}{8\pi}-\frac{\eta}{16\pi}) \ln(\frac{\sqrt{\alpha}}{2\pi})+\frac{\eta}{128\pi}-\frac{\eta^{2}}{512\pi\alpha}-\frac{17\alpha}{64\pi} \right]~.
\label{014}
\end{eqnarray}
In terms of the area of the black hole horizon $A$, eq.\eqref{eq18} can be recast in the form 
\begin{eqnarray}
S & = & \frac{A}{4l_{p}^{2}}+\left (\frac{\eta}{32\pi}-\frac{\alpha}{16\pi}\right )\ln (\frac{A}{4 \pi l_{p}^{2}})+\frac{\alpha^{2}}{64\pi^{2} }(\frac{l_{p}^{2}}{A})-\frac{\alpha \eta}{128\pi^{2} }(\frac{l_{p}^{2}}{A})+\frac{\eta^{2}}{512\pi^{2} }(\frac{l_{p}^{2}}{A}) + S_{0}  \nonumber \\ && + \mathcal{O}(\alpha^{3},\eta \alpha^{2}, \eta^2 \alpha, \eta^{3})~.
\label{eq19}
\end{eqnarray} 
Interestingly, for $\alpha = 0$, we can obtain the exact expression of entropy for higher dimensional RG inspired Schwarzschild black hole (using the expression of temperature in eq.\eqref{eq15})
\begin{eqnarray}
S = \frac{(d-2)\pi^{\frac{d-3}{2}}E_{p}^{(d-3)}\sqrt{\eta}r_{+}^{(d-3)}}{8\Gamma(\frac{d-1}{2})}{}_2F_{1}\left (-\frac{1}{2},\frac{(d-3)}{2},\frac{(d-1)}{2}; -\frac{16 \pi^{2}r_{+}^{2}E_{p}^{2}}{(d-3)^{2} \eta} \right )
\label{eq20}
\end{eqnarray}
 \noindent where ${}_2F_1$ is the Gauss hypergeometric function. This is a new result that we obtain from our analysis. For $d=4$, the above expression simplifies to
\begin{eqnarray}
S = \frac{r_{+}E_{p}}{4}\sqrt{16 \pi^{2}E_{p}^{2}r_{+}^{2}+\eta}+\frac{\eta }{16\pi}\sinh ^{-1}\left(\frac{4\pi E_{p}r_{+}}{\sqrt{\eta}}\right )
\label{eq21}
\end{eqnarray}
which agrees with that in the literature \cite{wk}. 
Note that the exact expression for the entropy for $d=4$ with different values of $n$ and the usual dispersion relation 
has been reported earlier in \cite{sa}. In this paper, we have found an exact expression of the entropy for $n=2$ in $d$-dimensions incorporating the MDR.

\noindent For $d=5$, $\alpha\neq 0$, we get the entropy to be
\begin{eqnarray}
S & = & \frac{\pi^{2}r_{+}^{3}E_{p}^{3}}{2}-\frac{3 }{16}(2\alpha-\eta)r_{+}E_{p}+\frac{3(8 \alpha^{2}-4 \alpha \eta + \eta^{2})}{256 \pi^2 r_{+}E_{p}} -\frac{(56 \alpha^{2} + 36 \alpha \eta +3\eta^2)}{256 \pi^2 \sqrt{\alpha}} \nonumber \\ && + \mathcal{O}(\alpha^{3},\eta \alpha^{2}, \eta^2 \alpha, \eta^{3})~.
\label{eq22}
\end{eqnarray}

\noindent In terms of the horizon area $A$, the expression for the entropy in $d=5$ takes the form
\begin{eqnarray}
S &=& \frac{A}{4l_{p}^{3}}-\frac{3\alpha}{(256 \pi^{2})^{\frac{1}{3}}}\left (\frac{A}{4l_{p}^{3}} \right )^{\frac{1}{3}}+\frac{3\eta}{(2048 \pi^{2})^{\frac{1}{3}}}\left (\frac{A}{4l_{p}^{3}} \right )^{\frac{1}{3}}+\frac{3\alpha^{2}}{16 (2\pi)^{\frac{4}{3}}}\left ( \frac{4l_{p}^{3}}{A} \right )^{\frac{1}{3}} -\frac{3\eta \alpha}{32(2\pi)^{\frac{4}{3}}}\left ( \frac{4l_{p}^{3}}{A} \right )^{\frac{1}{3}} \nonumber \\ && +\frac{3\eta^{2}}{128(2\pi)^{\frac{4}{3}}}\left ( \frac{4l_{p}^{3}}{A} \right )^{\frac{1}{3}} + S_{0} + \mathcal{O}(\alpha^{3},\eta \alpha^{2}, \eta^2 \alpha, \eta^{3}) ~.
\label{eq23}
\end{eqnarray}


\section{Free energy and phase transition}
In this section, we move on to study phase transition in a higher dimensional Schwarzschild black hole in RG incorporating the effects of the GUP. To proceed, we place the black hole inside a finite concentric spherical cavity whose radius is larger than that of the black hole horizon radius. The temperature is then fixed on the surface of the cavity and is said to be the local temperature $ T_{loc} $ as seen by a local observer. This reads \cite{tm,jw}
\begin{eqnarray}
T_{loc}=\frac{T}{\sqrt{-g_{00}}} =\frac{\frac{2 \pi r_{+} E_p^2}{ (d-3) \alpha}\left ( 1 - \sqrt{1 - \frac{(d-3)^2 \alpha }{4 \pi^2 r_{+}^{2} E_{p}^{2}}} \right )}{\sqrt{1 + \frac{4 \pi^2 r_{+}^{2}E_{p}^{2}\eta }{(d-3)^2 \alpha ^{2}}\left( 1 - \sqrt{1 - \frac{(d-3)^2 \alpha}{4 \pi^2 r_{+}^{2}E_{p}^{2}}}\right )^{2}}\sqrt{1-(\frac{r_{+}}{r})^{d-3}}}~.
\label{eq24}
\end{eqnarray} 
In the limit $\alpha \rightarrow 0$, the above expression takes the form
\begin{eqnarray}
T_{loc}=\frac{(d-3) E_p}{\sqrt{(16 \pi^2 E_{p}^{2}r_{+}^{2}+ (d-3)^2\eta )}\sqrt{1-(\frac{r_{+}}{r})^{d-3}}}~.
\label{tloc}
\end{eqnarray}
\noindent In Fig. \ref{fig:fig2}, we plot $T_{loc}$ vs $r_{+}$ for $d=4,5$ with two different values of the GUP parameter $\alpha= 0, 0.3$. The rainbow parameter $\eta$ has been taken to be unity and $E_{p}$ has been set to $1$. In Fig.\ref{fig:sfig3}, one can observe the existence of a local maximum temperature $T_{M}$ near the origin at $r_{+}=r_{M}$ and a local minimum temperature $T_{m}$ at $r_{+}=r_{m}$ (for $\alpha = 0$ and $d=4$). However, for $d=5$ the local maximum temperature does not exist (see Fig.\ref{fig:sfig5} ). For a non-zero value of the GUP parameter $\alpha$, we observe that the local temperature saturates to a maximum value at a nonzero value of the radius of the black hole. We denote this temperature as $T_{r}$ at $r_{+}= (r_{+})_{rem} =\frac{(d-3)}{2 \pi}\sqrt{\alpha}l_{p}$.

\noindent Next, making use of the expression for the entropy \eqref{012} and the local temperature \eqref{eq24}, the first law of thermodynamics yields 
\begin{eqnarray}
E_{loc}&=&\int_{(r_{+})_{rem}}^{r_{+}}T_{loc}dS \nonumber\\ & = &\int_{(r_{+})_{rem}}^{r_{+}}\frac{T_{loc}}{T}dM \nonumber\\ & = & \frac{(d-2)\Omega_{d-2}r^{d-3}}{8\pi l_{p}^{d-2}}\left (\sqrt{1-(\frac{(r_{+})_{rem}}{r})^{d-3} }-\sqrt{1-(\frac{r_{+}}{r})^{d-3} } \right )
\label{eq25}
\end{eqnarray}
where we have used eq.(s)(\ref{eq9}, \ref{04}).
In the limit $\alpha \rightarrow 0$, $(r_{+})_{rem}=0$ and hence the above expression takes the form
\begin{eqnarray}
E_{loc}=\frac{(d-2)\Omega_{d-2}r^{d-3}}{8\pi l_{p}^{d-2}}\left (1-\sqrt{1-(\frac{r_{+}}{r})^{d-3} } \right )~.
\label{eq26}
\end{eqnarray}

\begin{figure}
\begin{subfigure}{.5\textwidth}
  \centering
  \includegraphics[width=  7.5cm, height = 5cm]{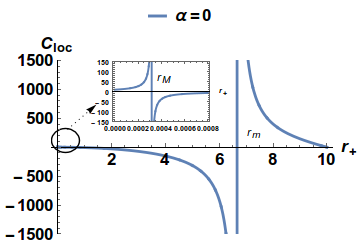}
  \caption{d=4}
  \label{fig:sfig7}
\end{subfigure}%
\begin{subfigure}{.5\textwidth}
	\centering
	\includegraphics[width=  7.5cm, height = 5cm]{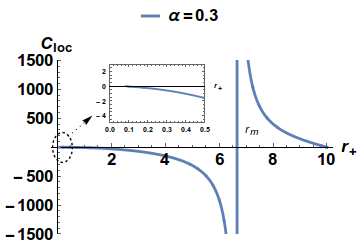}
	\caption{d=4}
	\label{fig:sfig8}
\end{subfigure}  \nonumber\\
\begin{subfigure}{.5\textwidth}
	\centering
	\includegraphics[width=  7.5cm, height = 5cm]{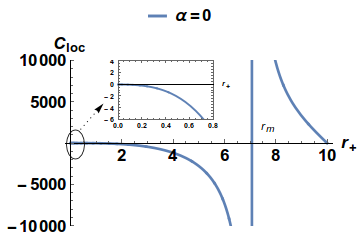}
	\caption{d=5}
	\label{fig:sfig9}
\end{subfigure}%
\begin{subfigure}{.3\textwidth}
  \centering
  \includegraphics[width= 7.5cm, height = 5cm]{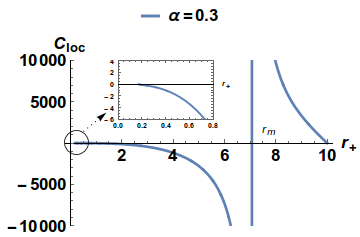}
  \caption{d=5}
  \label{fig:sfig10}
\end{subfigure}
\caption{ Heat capacity vs horizon radius.}
\label{fig:fig3}
\end{figure}

\begin{figure}
\begin{subfigure}{.5\textwidth}
  \centering
  \includegraphics[width=  7.5cm, height = 5.5cm]{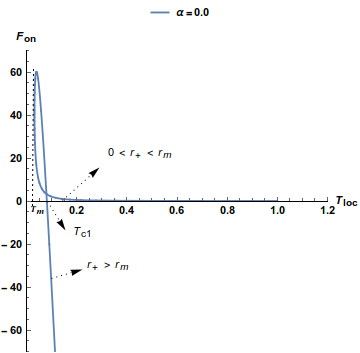}
  \caption{d=5}
  \label{fig:sfig11}
\end{subfigure}%
\begin{subfigure}{.5\textwidth}
  \centering
  \includegraphics[width= 7.5cm, height = 5.5cm]{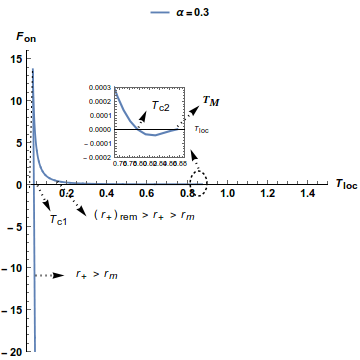}
  \caption{d=5}
  \label{fig:sfig12}
\end{subfigure}%
\caption{Variation of free energy vs. local temperature $T_{loc}$ in $d=5$ dimensions. The choice of the parameters are $\eta=1$, $r=10$ and $E_{p}=1$  with $\alpha=0.0,0.3 $.}
\label{fig:fig4}
\end{figure}

\noindent To investigate the thermodynamic stability of this black hole in RG, we now calculate the local heat capacity $C_{loc}$. This reads
\begin{eqnarray}
C_{loc}&=& \frac{\partial E_{loc}}{\partial T_{loc}} \nonumber\\ &
=& \frac{A(r_{+},r,\eta,\alpha,E_{p})}{B(r_{+},r,\eta,\alpha,E_{p})}
\label{eq27}
\end{eqnarray}
where 
\begin{eqnarray} 
A(r_{+},r,\eta,\alpha,E_{p})&=& -(d-2)(d-3) \alpha E_{p}^{d-2} \pi^{\frac{d-3}{2}}  r_{+}^{d+3}\left(1- (\frac{r_{+}}{r})^{d-3}\right) \sqrt{4 \pi^2- \frac{\alpha (d-3)^2}{E_{p}^2 r_{+}^2}} \nonumber\\ && \left(\alpha \eta (d-3)^2 - \alpha^2 (d-3)^2 +  4 \eta \pi E_{p}^2 r_{+}^2 \left(-2 \pi + \sqrt{4 \pi^2- \frac{\alpha (d-3)^2}{E_{p}^2 r_{+}^2}}\right) \right) \nonumber\\ && \sqrt{1+\frac{2 \eta E_{p}^2 \pi^2 r_{+}^2 \left(1- \sqrt{1- \frac{\alpha (d-3)^2}{4 \pi^2 E_{p}^2 r_{+}^2}}\right)^2}{\alpha^2 (d-3)^2}}
\label{eq28}
\end{eqnarray}

\begin{eqnarray}
B(r_{+},r,\eta,\alpha,E_{p})&=&\frac{4r_{+}^{2}}{\Gamma(\frac{d-1}{2})}\left[(d-3)^{4}\left(\frac{r_{+}}{r} \right)^{d}r^{3}\alpha^{2}(\alpha -\eta) + 32\eta \pi^{3}E_{p}^{4}r^{3}r_{+}^{4} \left(\frac{r_{+}}{r} \right)^{d} \right. \nonumber\\ && \left. \big(-2\pi +\sqrt{4\pi ^{2}-\frac{(d-3)^{2}\alpha }{r_{+}^{2}E_{p}^{2}}}\big) - 2(d-3)\alpha \pi E_{p}^{2}r_{+}^{2} \right. \nonumber\\ && \left. \left(4\pi \alpha r_{+}^{3} + 2\pi r^{3}( \frac{r_{+}}{r} )^{d} \left((d-5)\alpha - 5\eta (d-3)\right)  - \sqrt{4\pi ^{2}-\frac{(d-3)^{2}\alpha }{r_{+}^{2}E_{p}^{2}}} \right. \right. \nonumber\\ && \left. \left. \left( 2\alpha r_{+}^{3}+r^{3}(\frac{r_{+}}{r})^{d} \big( (d-5)\alpha -3\eta (d-3)\big)\right) \right) \right]~~.
\label{eq29}
\end{eqnarray}

\noindent In Fig.\ref{fig:fig3}, the specific heat is plotted as a function of $ r_{+} $. For $d=4$ and $\alpha=0$, it shows three qualitatively different regions. For $d=5$, there are two different regions in contrast to $d=4$ \cite{wk}. It is well known that the black hole is stable if the specific heat is positive while it radiates if it has negative heat capacity. For $d=4$ and for $ r_{+} > r_{m} $ , the specific heat is positive which implies that the black hole is stable. This is called the large stable black hole (LSB). For $ r_{M} > r_{+} > r_{m} $, the black hole has negative heat capacity, hence the black hole is unstable. This is called small unstable black hole (SUB). In the region $ 0 < r_{+} < r_{M} $, the heat capacity is positive. This type of black hole called tiny stable black hole (TSB). For $d=5$ or any higher dimension, the region $0 < r_{+} < r_{M} $ is absent. Hence only two black hole phases (LSB for $ r_{+} > r_{m} $ and SUB for $ 0 > r_{+} > r_{m} $) exist for any higher dimension $d \geq 5$.

\noindent For a non-zero value of the GUP parameter $\alpha$, we observe that the heat capacity  vanishes at a non-zero value of $r_{+}$ implying the existence of black hole remnants. Further, in this case, there are only two regions, namely, the LSB (for $ r_{+} > r_{m} $) and the SUB (for $ (r_{+})_{rem} > r_{+} > r_{m} $).

\noindent With the above results in hand, we are now in a position to study phase transition of the black hole. For that we calculate the on-shell free energy of the higher dimensional  Schwarzschild black hole in RG incorporating the effects of the GUP \cite{jw,cai}. This reads 
\begin{equation}
F_{on}=E_{loc}-T_{loc}S
\label{eq30}
\end{equation}
where $E_{loc}$, $T_{loc}$ and $S$ are given in eq.(s)(\ref{eq25}, \ref{eq24}, \ref{eq22})respectively.

\noindent In Fig.\ref{fig:fig4}, we have shown  $F_{on}$ curves with $T_{loc}$ for different values of $\alpha$ for $d=5$. We shall first discuss the case where the GUP parameter $\alpha=0$. Before we proceed, we would like to define the hot flat space (HFS) as one in which $F_{on}=0$ for any arbitrary $T_{loc}$. This happens when $r_{+} \rightarrow 0$ where $E_{loc}=0$, $S=0$ as seen from eq.(s) (\ref{eq26}, \ref{eq20}) respectively. Hence the HFS free energy $F_{on}^{HFS}=0$. When the temperature of the black hole ($T$) is less than the critical temperature ($T_{c1}$), that is, $T_{m} <T < T_{c1}$, the on-shell free energy of the black hole obeys the relation  $F_{on}^{HFS}<F_{on}^{SUB}<F_{on}^{LSB}$ where $F_{on}^{SUB}$ and $F_{on}^{LSB}$ are the free energies of the SUB and the LSB. It implies that the HFS is more probable than both the LSB and SUB in this temperature region. This feature is similar to that observed in the literature for $d=4$. However, in $d=5$, there are three possible states namely, the LSB, SUB and HFS and the decay of the black hole state takes place from the LSB to SUB and then from SUB to HFS. This is in contrast to the $d=4$ case, where the decay takes place from SUB to HFS \cite{wk}. For $T>T_{c1}$, we have $F_{on}^{SUB}>F_{on}^{HFS}>F_{on}^{LSB}$. Therefore, when the temperature of the back hole is greater than the critical temperature $T_{c1}$, the SUB decays to the HFS which eventually decays into the LSB. This type of phase transition is called the Hawking-Page phase transition which occurs at the critical point $T_{c1}$.

\noindent Now in the presence of the GUP parameter ($\alpha \neq 0 $), we have to redefine the hot curved space (HCS) \cite{kim} instead of the HFS due to the presence of the black hole remnants $(r_{+})_{rem}$. Here also, the free energy $F_{on}^{HCS}=0$ for all $T_{loc}$ as $E_{loc}=0$, $S=0$ as seen from eq.(s) (\ref{eq25}, \ref{eq22}). Here, we observe that there are two Hawking-Page type critical temperatures, namely, $T_{c1}$ and $T_{c2}$ in contrast to the case $\alpha = 0$ case. In the temperature region between $T_{c1} <T<T_{c2}$, the behavior of black hole phase transition is analogous to the scenario for $\alpha = 0$. For $T_{M}>T>T_{c2}$, the free energy satisfies $F_{on}^{HCS}>F_{on}^{SUB}>F_{on}^{LSB}$. The above relation states that the on-shell free energy of the SUB is still larger than the LSB. Therefore, the SUB undergoes a tunneling and eventually decays into the LSB. Finally for $T>T_{m}$, there are only two states of the black hole possible, namely, the HCS and the LSB and the decay occurs from the HCS to the LSB. The features obtained for $d=5$ are similar to those observed in $d=4$.


\section{Conclusions}
We summarize our findings now. In this paper we investigate the effect of the generalized uncertainty principle in the thermodynamics of the higher dimensional rainbow inspired Schwarzchild black hole using the modified dispersion relation. We first compute the mass-temperature relation for the black hole in $d$-dimensions which incorporates the effects of both the generalized uncertainty principle and rainbow gravity. The computation of the specific heat of the black hole from the mass-temperature relationship reveals the existence of remnants if the generalized uncertainty principle is taken into consideration. This is consistent with previous findings in the literature \cite{sg}. Interestingly, the remnant mass shows that there is no rainbow gravity dependency and it depends only on the generalized uncertainty principle parameter. However, remnants are still there in the absence of the generalized uncertainty relation if ordinary dispersion relation is used in the rainbow gravity framework. We then compute the entropy of the black hole in $d=4$ and $d=5$ dimensions. The area theorem is recovered along with the logarithmic correction in $d=4$ dimensions. In $d=5$ dimensions, there is however no logarithmic correction. An interesting observation that we made in our paper is that the results for the entropy are exact in any dimension in the absence of the GUP parameter $\alpha$. This is a new finding in this paper. We then move on to investigate the phase transition and thermodynamic stability of the black hole introducing the concept of a local observer. 
To do this, we first calculate the local specific heat and the on-shell free energy of the higher dimensional  Schwarzschild black hole in rainbow gravity incorporating the effects of the generalized uncertainty principle. It has been observed earlier that for $d=4$ and in the absence of the generalized uncertainty principle (that is $\alpha= 0$), there are three possible states of the black hole, namely, the large stable black hole, small unstable black hole and the tiny stable black hole. On the other hand, in $d=5$ and for $\alpha=0$, there are only two states of the black hole possible, namely, the large stable black hole and the small unstable black hole inspite of the absence of black hole remnants in this case. 
Here we conclude that there is only one phase transition of the black hole at $T_{c1}$ below which the hot flat surface becomes more probable than the small unstable black hole and the large stable black hole. 
The presence of the generalized uncertainty principle introduces the black hole remnants which in turn implies that the tiny stable black hole state do not exist both in $d=4$ and $d=5$ dimensions. However, here we get an additional critical temperature apart from the well known critical temperature in Hawking-Page phase transition.

\section*{Acknowledgments} RM would like to thank DST-INSPIRE, Govt. of India for financial support. RM would also like to thank Anant Vijay Varma of IISER-Kolkata for helping in Mathematica. SB would also like to thank Government of West Bengal for financial support. SG acknowledges the support by DST SERB under Start Up Research Grant (Young Scientist), File No.YSS/2014/000180. SG also acknowledges IUCAA, Pune for the Visiting Associateship.\\

\noindent \textbf{Author Contribution Statement :} All authors have contributed equally.


\end{document}